\documentclass[twocolumn,secnumarabic,amssymb, nobibnotes, aps, pre, superscriptaddress]{revtex4-1}

\newcommand{\rv}{{\bm r}}

\newcommand{\vv}{{\bm v}}

\newcommand{\lambdav}{\bm\lambda}

\usepackage{amsmath,amssymb}
\usepackage{graphicx} 
\usepackage{longtable}

\begin{document}


\title{\textbf{Assembly of hard spheres in a cylinder: a computational and experimental study}}


\author{Lin Fu}
\affiliation{NSF Research Triangle Materials Research Science and Engineering Center, Duke University, Durham, NC 27708, USA, E-mail: gabriel.lopez@duke.edu, patrick.charbonneau@duke.edu}
\affiliation{Department of Chemistry, Duke University, Durham, NC 27708, USA.}
\author{Ce Bian}
\affiliation{NSF Research Triangle Materials Research Science and Engineering Center, Duke University, Durham, NC 27708, USA, E-mail: gabriel.lopez@duke.edu, patrick.charbonneau@duke.edu}
\affiliation{School of Physics, Shandong University, Jinan, Shandong, 250100, China}
\author{C. Wyatt Shields IV}
\affiliation{NSF Research Triangle Materials Research Science and Engineering Center, Duke University, Durham, NC 27708, USA, E-mail: gabriel.lopez@duke.edu, patrick.charbonneau@duke.edu}
\affiliation{Department of Mechanical Engineering and Materials Science, Duke University, Durham, NC 27708, USA.}
\author{Daniela F. Cruz}
\affiliation{NSF Research Triangle Materials Research Science and Engineering Center, Duke University, Durham, NC 27708, USA, E-mail: gabriel.lopez@duke.edu, patrick.charbonneau@duke.edu}
\affiliation{Department of Biomedical Engineering, Duke University, Durham, NC 27708, USA.}
\author{Gabriel P. L\'opez}
\affiliation{NSF Research Triangle Materials Research Science and Engineering Center, Duke University, Durham, NC 27708, USA, E-mail: gabriel.lopez@duke.edu, patrick.charbonneau@duke.edu}
\affiliation{Department of Mechanical Engineering and Materials Science, Duke University, Durham, NC 27708, USA.}
\affiliation{Department of Biomedical Engineering, Duke University, Durham, NC 27708, USA.}
\author{Patrick Charbonneau}
\affiliation{NSF Research Triangle Materials Research Science and Engineering Center, Duke University, Durham, NC 27708, USA, E-mail: gabriel.lopez@duke.edu, patrick.charbonneau@duke.edu}
\affiliation{Department of Chemistry, Duke University, Durham, NC 27708, USA.}
\affiliation{Department of Physics, Duke University, Durham, NC 27708, USA.}


\date{\today}

\begin{abstract}
Hard spheres are an important benchmark of our understanding of natural and synthetic systems. In this work, colloidal experiments and Monte Carlo simulations examine the equilibrium  and out-of-equilibrium assembly of hard spheres of diameter $\sigma$ within cylinders of diameter $\sigma\leq D\leq 2.82\sigma$. Although in such a system phase transitions formally do not exist, marked structural crossovers are observed. In simulations, we find that the resulting pressure-diameter structural diagram echoes the densest packing sequence obtained at infinite pressure in this range of $D$. We also observe that the out-of-equilibrium self-assembly depends on the compression rate. Slow compression approximates equilibrium results, while fast compression can skip intermediate structures. Crossovers for which no continuous line-slip exists are found to be dynamically unfavorable, which is the source of this difference. Results from colloidal sedimentation experiments at high P\'eclet number are found to be consistent with the results of fast compressions, as long as appropriate boundary conditions are used. The similitude between compression and sedimentation results suggests that the assembly pathway does not here sensitively depend on the nature of the out-of-equilibrium dynamics.
\end{abstract}

\pacs{}

\maketitle

\section{Introduction}

Dense packings of hard objects have long fascinated mathematicians and condensed matter physicists~\cite{conway2013sphere}. The resulting structures are indeed both elegant and useful~\cite{frenkel2010viewpoint,donev2004improving,henzie2012self,boles2016self}. Although it typically receives less attention, their assembly pathways can be just as important~\cite{damasceno2012predictive,whitelam2014viewpoint}.  While failure to assemble optimal structures often results in gel or glass formation, ordered suboptimal outcomes, including limit periodic structures~\cite{marcoux2014emergence} and quasi-crystals~\cite{haji2009disordered}, are also observed.  Because formulating a general theory of out-of-equilibrium dynamics is challenging, insights into self-assembly are often first obtained through numerical simulations. Colloidal suspensions, in which particle interactions can be tuned~\cite{ivlev2012complex,li2011colloidal,mewis2012colloidal} and their real-space positions monitored through microscopy, often complement and enrich numerical results. Hard-sphere--like colloids, for instance, have been used to examine crystallization, glass formation~\cite{cipelletti2000universal,pusey1987observation,weeks2000three} and assembly under external fields~\cite{ristenpart2003electrically,yang2015phase,owens2016highly}. 

Conceptually, the dynamical accessibility of structures is often couched in terms of geometrical frustration~\cite{manoharan2015colloidal,sadoc2006geometrical,nelson2002defects,grason2016perspective}. If the locally preferred organization in the disordered phase is compatible with the globally optimal structure, the system is deemed unfrustrated and assembly is presumed facile. A classic example is a system of two-dimensional hard disks. Locally, the disks spontaneously form equilateral triangles, which can easily organize into the globally optimal triangular lattice. By contrast, a mismatch between the two types of order, i.e., a geometrically frustrated system, places hurdles on the path to assembly. Three-dimensional spheres, for instance, locally form perfect tetrahedra, but their optimal packings (face-centered cubic, hexagonal close-packed, etc.) require the inclusion of geometrical features other than tetrahedra, thus giving rise to a rich set of assembly intermediates~\cite{auer2001prediction,schilling2010precursor,schope2006two,van1997template,cheng2001crystallization}. A counterpart to this effect is that a suboptimal structure may sometime form more efficiently and reliably than the optimal one, simply because the assembly pathway to the former is less geometrically frustrated than the latter. 

Another family of mechanisms for facile assembly are transformations that allow the continuous deformation of one ordered structure into another. Martensitic transformations, which are commonly observed in steel and other metallic alloys~\cite{nishiyama2012martensitic,khachaturyan2013theory} and have also been reported in various colloidal systems~\cite{yethiraj2004nature,nojd2013electric,mohanty2015multiple,yang2015phase}, are prime examples of such processes~\cite{khachaturyan2013theory}. They are straightforward and reversible because their course does not require long-range particle rearrangements, i.e., they are are diffusionless.
 
Although these and other elementary self-assembly principles are often separately invoked, they are not so commonly at play within a same family of systems~\cite{chen2014}. In this respect, packings of hard spheres of diameter $\sigma$ in a cylinder are particularly interesting. Different assembly mechanisms are necessarily involved because a rich variety of optimal structures emerge upon making small changes to the cylinder diameter, $D$. From $D=\sigma$ to $D=4\sigma$, for instance, optimal structures go from being a straight chain to simple zigzags to phyllotactic-like chiral helices~\cite{pickett2000,mughal2012}, to quasiperiodic and complex packings with rattling particles~\cite{fu2016}. Interestingly, in these systems particle-particle interactions are short-ranged and the geometry is quasi-one-dimensional, hence any order that develops can only have a finite spatial extent at finite temperature and pressure. Because all structural changes must thus be formally continuous~\cite{van1950integrale,ruelle1969rigorous,lieb2013mathematical}, the order of the transition cannot directly affect the ease of assembly.  
Beyond the trivial regime $D\leq\sqrt{3}/2\sigma$, in which the simple transfer-matrix treatment is exact~\cite{mon2000hard,kamenetskiy2004equation,varga2011structural,gurin2013pair,boublik2010hard,kofke1993hard,kim2011thermodynamics} (a similar treatment should be possible for $\sqrt{3}/2\sigma< D<2\sigma$ by including second-nearest neighbor interactions, as in Ref.~\cite{gurin2015beyond}, but has yet to be considered for this geometry), prior work on this question, whether experimental~\cite{lohr2010helical,kumzerov1990,jiang2013helical} or computational~\cite{koga2006,duran2009ordering,huang2010direct,gordillo2006freezing,huang2009characterization,yue2011spontaneous}, is relatively sparse.  Accurate knowledge of even the densest packings was indeed missing until fairly recently~\cite{mughal2012,fu2016}. 
 From these pioneering studies, we nonetheless know that the equilibrium behavior qualitatively differs for cylinders with $D<2\sigma$ and $D\geq 2\sigma$. This distinction is likely related to particles being able to pass one another in the latter, but not in the former.

In this work, we perform a detailed computational study of hard spheres under different degrees of cylindrical confinement, $\sigma\leq D\le2.82\sigma$. We obtain the equilibrium structural diagram using specialized Monte Carlo simulations, examine the assembly dynamics and the accessibility of dense packings by compression, and compare the results with an experimental setup in which micron-sized particles sediment within cylindrical pores. In Sections~\ref{sec:comMeth} and ~\ref{sec:expMeth}, we describe the computational and experimental methods, respectively, and in Section ~\ref{sec:results} we present and discuss equilibrium and out-of-equilibrium simulation results. This analysis is then used to interpret the sedimentation experiments.

\section{Computational Method}
\label{sec:comMeth}
We consider a system of $N$ hard spheres of diameter $\sigma$, confined within a hard cylinder of fixed diameter $D$ and variable height $\lambda_{\rm z}$. The system is under periodic boundary conditions along $z$ with an angular twist at the boundary. In this setup, a periodic packing is described by
\begin{equation}
\rv_{i\alpha}=\rv_i+n_\alpha\lambdav,
\end{equation}
where $\rv_{i\alpha}$ is the position of the $i$th particle within the $\alpha$th unit cell, $n_{\alpha}\in\mathbb{Z}$, and $\lambdav=(\lambda_{\mathrm{r}},\lambda_{\mathrm{\theta}} ,\lambda_{\mathrm{z}})$ is the lattice vector with subscripts $\mathrm{r}$, $\mathrm{\theta}$ and $\mathrm{z}$ denoting the axial, radial and angular components of cylindrical coordinates, respectively. Hence, $\lambda_{\mathrm{\theta}}$ is the twist angle at the periodic boundary and $\lambda_{\mathrm{z}}$ is the unit cell height. For all the systems studied here, $\lambda_{\mathrm{r}}=0$.

Monte Carlo (MC) simulations are run in the isothermal-isobaric (constant-$NPT$)  ensemble. Pressure $P$ is kept constant by standard logarithmically-sampled volume moves on $z$~\cite{frenkel2002}, while temperature, $T$, is a trivial scaling factor for hard interactions. Particles evolve through random local displacements that are accepted following the Metropolis criterion, and so is the twist angle $\lambda_{\mathrm{\theta}}$. The step size of the various MC moves is tuned to ensure an acceptance ratio between 30\% to 40\%. The remaining simulation parameters are chosen differently, depending on whether equilibrium or compression simulations are performed, as described below.

\subsection{Equilibrium Simulations}
\label{sec:ECMC}
Confinement hinders structural relaxation, especially at high pressures. In order to accelerate equilibration, we adapt the event-chain Monte Carlo (ECMC) scheme~\cite{bernard2009event} to the cylinder geometry, which accelerates sampling by up to four orders of magnitude at the highest pressure studied. ECMC is a rejection-free sampling scheme that allows an arbitrarily long chain of particles to be displaced in a single move, while preserving algorithmic balance. In general, one first randomly selects a particle and a vector $\vv$ for displacing it. After colliding with another particle, the initial one is immobilized and the second particle is displaced until it itself collides, and so on. The procedure terminates once one of the components of the sum of individual displacements reaches a given $l$. Choosing $l$ to be comparable to the size of the simulation box ensures that a collective displacement is achieved.

In order to adapt this scheme to cylindrical confinement, we restrict $\vv$ to have $|v_\mathrm{r}/v_\mathrm{z}|\le0.5$, so that particle displacements lie mainly along $z$; $l=\lambda_{\mathrm{z}}$ is also chosen along $z$; and if a particle collides with the wall of the cylinder before reaching another particle, it is reflected by the plane tangent to the collision point. The detailed procedure is as follows.
\begin{enumerate}
\item Randomly select particle $i\in[1,N]$ and displacement vector $\vv$, such that $v_\mathrm{r}/v_{\mathrm{z}}\in[-0.5,0.5]$ and $v_{\mathrm{\theta}}\in[0,2\pi)$.

\item For all $j\ne i$, calculate the distance $d_{ij}$ that $i$ would travel along direction $\vv$ before hitting $j$, and find $j^{\mathrm{min}}=\underset{j}{\operatorname{argmin}}\ d_{ij}$. (A cell list is used to accelerate this step.) Calculate also the distance $d_{\mathrm{wall}}$ that $i$ would travel before hitting the cylinder wall.

\item Denoting the $z$ component of the cumulative displacements $z_{\rm cum}$, calculate $d_{\rm move}$=min($d_{\rm wall}$, $d_{ij^{\mathrm{min}}}$, $l-z_{\rm cum}$).
\begin{itemize}
\item If $d_{\rm move}=d_{\rm wall}$, move particle $i$ to the wall by $d_{\rm wall}$, reflect $\vv$, and then go back to step 2;
\item If $d_{\rm move}=d_{ij^{\rm min}}$, move particle $i$ up to its collision point with $j^{\rm min}$ by $d_{ij^{\mathrm{min}}}$, replace $i$ by $j^{\mathrm{min}}$, and then go back to step 2;
\item If $d_{\rm move}=l - z_{\rm cum}$, move particle $i$ by $l-z_{\rm cum}$ and terminate.
\end{itemize}
\end{enumerate}

In large systems, relatively small volume moves are conducted as otherwise they are rejected with high probability. In order to further improve their computational efficiency, we keep track of the minimum collision distance between particles in the $z$ direction, $z_{\mathrm{max}}\equiv\max\ z_{ij}$, $i\neq j,$ where $z_{ij}= \sqrt{\sigma^2 - r_{ij}^2 + (r_{i\mathrm{z}} - r_{j\mathrm{z}})^2} / \Delta z_{ij}$ with $\Delta z_{ij}=(r_{i\mathrm{z}} - r_{j\mathrm{z}})$. For a system without overlap, we thus obtain:
\begin{equation}
 z_{ij}^2=\frac{\sigma^2 - (r_{ij}^2 - \Delta z_{ij}^2)}{\Delta z_{ij}^2}\le1,\ \ \ \ \forall i,j.\label{eqn:zmax}
\end{equation}
Because the numerator in Eqn.~\eqref{eqn:zmax} does not contain $\Delta z_{ij}$, each volume move rescales $1/z_{\mathrm{max}}$ linearly with $\lambda_{\mathrm{z}}$, and as long as $z_{\mathrm{max}}\le 1$ no overlap occurs.
This simple bookkeeping keeps the computational cost of volume moves of $\mathcal{O}(1)$. For systems initially prepared far from equilibrium, the frequency of volume moves should not be too high in order to maintain an efficient balance with particle moves. We find that having each MC average $N/60$ volume, one ECMC, $N$ particle, and 10 twist angle moves ensures fast and reliable equilibration. 

As a check of equilibration, simulations are initialized from at least two different conditions: (i) disordered configurations at a low packing fraction, $\eta = 0.05$; and (ii) lattice configurations of the densest packings for the cylinder diameter considered. Around the structural crossovers, lattice configurations of intermediate order types are additionally used as starting points. In all cases, equilibration is run over $1.0\times10^7$ MC steps before observables are sampled over $5.0\times10^6$ MC steps. The final system is deemed equilibrated if the different preparation protocols provide numerically consistent results.

We sample $\sigma\leq D\leq 2.82\sigma$ with a typical resolution of $0.33\sigma$. Because the structural correlation length, $\xi$, grows significantly as pressure increases -- especially for achiral structures -- a sufficiently large $N$ is used to prevent self-correlations with the periodic image. We find $N=3000$ to suffice over the $P$ and $D$ regime sampled, which keeps the equilibrium unit cell height reasonably large, i.e., $\lambda_{\mathrm{z}}\gtrsim 15\xi$. Additional trial runs with $N=5000$ and 10000 for certain $D$ validate this choice.

\subsection{Assembly Simulations}

The impact of the compression rate on assembly is studied by MC pressure annealing. Simulations all start from a disordered, low-density structure at $\eta = 0.05$. The unitless pressure, $P^*_{\rm init}\equiv\beta P \sigma^3=1.0$ is then ramped up to $P^*_{\mathrm{max}} = 50$, in steady increments after each MC cycle. The compression rate,  $\gamma = (P^*_{\mathrm{max}}-P^*_{\mathrm{init}})/t_{\mathrm{MC}}$, where $t_{\mathrm{MC}}$ is the number of MC cycles.
We consider the assembly of systems under fast and slow compressions, $\gamma=4.90\times10^{-4}$ and $\gamma=2.45\times10^{-7}$, respectively, over the full range of cylinder diameters studied above. We also consider the impact of intermediate rates $1.63\times10^{-5}\leq\gamma\leq2.45\times10^{-7}$ for $D=2.20\sigma$. To explore the role of the cylinder boundary in experiments, simulations with a hard bottom under fast compression are additionally performed.

For assembly simulations, we use $N=60$, which approximates the experimental system size, and eliminate ECMC moves because they do not correspond to a physical (local) dynamics. Each MC cycle thus averages one volume, $N$ particle, and 10 twist angle moves.  We explore the diameter range $2.00\sigma\le D\le 2.82\sigma$ with a resolution of 0.01$\sigma$, and consider 20 assembly simulations for each $D$ because different structures can assemble for a same $D$. Below  $2\sigma$, assembly is as trivial as the optimal structure. Above 2.82$\sigma$, structures are found with inner particles that do not touch the cylinder wall, which is a structural complexity beyond scope of this work. For $D=2.2\sigma$, which is examined in finer detail, 300 simulations are performed for each compression rate.

\subsection{Correlation Length Calculation}
\begin{figure}[h]
\centering
  \includegraphics[width=1.0\columnwidth]{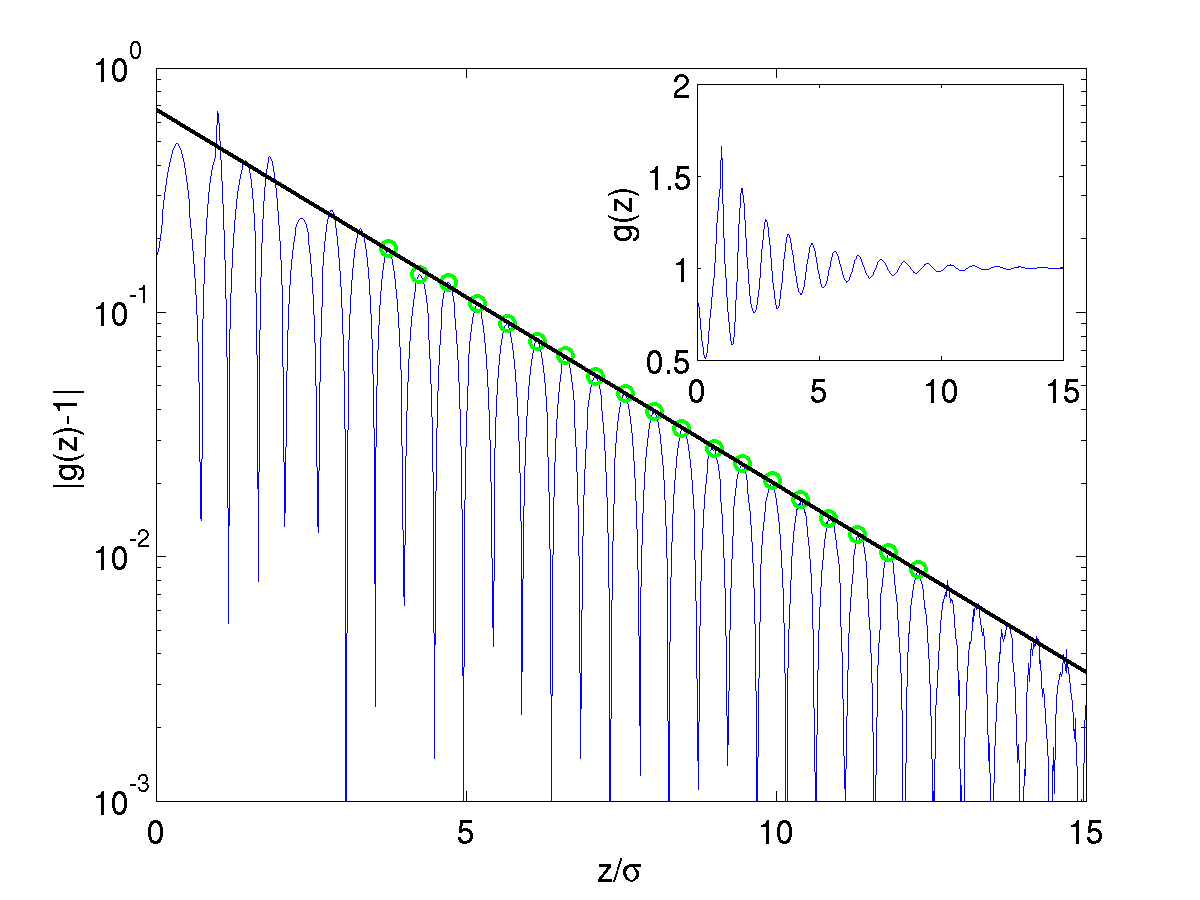}
  \caption{Radial decay of the axial component of the pair correlation function at $P^*=6.0$ for $D=2.40\sigma$. Green circles denote the points used to fit Eq.~\eqref{eqn:semilog}, and the straight line shows the results for a decay length $\xi=2.83\sigma$. (Inset) Linear-scale oscillations of $g(z)$.}
    \label{fig:semilog}
\end{figure}

The equilibrium structural correlation length, $\xi$, characterizes the spatial decay of the axial component of the pair correlation function, $g(\rv)$, 
\begin{equation}
g(z)=\frac{1}{\rho}\left\langle\frac{1}{N}\sum_{i<j} \delta(|r_{iz}-r_{jz}|-z)\right\rangle,
\end{equation}
where $\rho=N/V_{\mathrm{in}}$, and $V_{\mathrm{in}}=\pi[(D-\sigma)/2]^2\lambda_{\mathrm{z}}$. This function effectively projects particle positions onto  the $z$ axis of the cylinder, before computing the standard pair correlation function. The correlation length is then obtained from a logarithmic fit of the exponential decay of the peaks of $g(z)$, 
\begin{equation}
g(z)=1+A\mathrm{e}^{-z/\xi},
\label{eqn:semilog}
\end{equation}
where the prefactor $A$ is fitted along $\xi$. Note that for the systems studied here, the first few peaks have to be discarded from the analysis because preasymptotic corrections can hide the exponential decay at short length scales, while peaks at large $z$ are discarded once they become comparable to the statistical noise. An exponential decay over more than two decades is then typically obtained (see Fig.~\ref{fig:semilog}).

\subsection{Structural Parameter}
\label{sec:OP}

In the regime $2.00\sigma<D\le2.82\sigma$, configurations can be described as a rolled two-dimensional sheet of (nearly) close-packed hard spheres. We devise a structural parameter, $\theta_6$, that measures the orientation of a hexagon on this sheet relative to the $x$-$y$ plane, i.e., the plane perpendicular to the $z$ axis of the cylinder:
\begin{equation}
\theta_6 \equiv \left\langle\frac{\mathrm{arg}[\sum_{k = 1}^{N_{\mathrm{nn}}}\exp(i 6\theta_{jk})]}{6}\right\rangle,
\end{equation}
where $N_{\mathrm{nn}}$ is the number of nearest neighbors within $1.3\sigma$ of particle $j$, and $\theta_{jk}$ is the bond angle between the $x$-$y$ plane and the bond connecting particle $j$ to its nearest neighbor $k$. Note that although no complete hexagon can be found when unrolling the structure of narrow cylinders, $\theta_6$ nonetheless captures structural differences between different packings. Changes to $\theta_6$ are thus more revealing than its absolute value. For disordered structures, however, $\theta_6$ presents no clear signature, and is thus not reported. 

To detect ordering at low pressures, we use instead the two-dimensional hexatic order parameter:
\begin{equation}
\Phi_6=\left\langle\frac{\left|\sum_{k=1}^{N_{\mathrm{nn}}}\mathrm{exp}(i6\theta_{jk})\right|}{N_{\mathrm{nn}}}\right\rangle.
\end{equation}
For $D>2.00\sigma$, an inflection point in the equations of state is observed for $P^*\approx8$, which hints at the existence of a crossover between ordered and disordered configurations. We find $\Phi_6=0.7$ to be a geometrical threshold comparable to this pressure. For the sake of convenience we thus consider structures to be ordered when $\Phi_6\ge0.7$.

\subsection{Structural Notation}

The patterns on the unrolled sheets can be described using a phyllotactic notation: $(l,m,n)$ with $l=m+n$ and $m\ge n$, where $l$, $m$ and $n$ are then number of helices along the three possible helical directions. In particular, $(l,l,0)$ corresponds to piles of staggered rings~\cite{mughal2012,fu2016}. To identify the dominant helical symmetry without visual inspection and detect the onset of structural changes, we calculate typical $\theta_6$ for various structures (Table~\ref{tbl:theta_6}). Note that $\theta_6$ differs for the two isomers of a chiral structure, but their sum is always $\pi/3$.

\begin{table}[h]
\small
  \caption{Structural parameter $\theta_6$ for perfect helices, rounded to the fourth decimal. For chiral structures, only the smaller of the two values is reported (see text).}
  \label{tbl:theta_6}
  \begin{tabular}{@{\extracolsep{\fill}}lllll}
    \hline
    Structure & $\theta_6$/rad & Structure & $\theta_6$/rad\\
    \hline
    (3,2,1) & $\mathrm{arctan}(\sqrt{3}/5)\approx0.3335$ & (4,4,0) & $\mathrm{arctan}(0)=0.0000$\\
    (3,3,0) & $\mathrm{arctan}(0)=0.0000$ & (5,3,2) & $\mathrm{arctan}(\sqrt{3}/4)\approx0.4086$\\
    (4,2,2) & $\mathrm{arctan}(\sqrt{3}/3)\approx0.5236$ & (5,4,1) & $\mathrm{arctan}(\sqrt{3}/9)\approx0.1901$\\
    (4,3,1) & $\mathrm{arctan}(\sqrt{3}/7)\approx0.2426$ & (6,3,3) & $\mathrm{arctan}(\sqrt{3}/3)\approx0.5236$\\
    \hline
  \end{tabular}
\end{table}

The unrolled sheet is decorated with perfect equilateral hexagons only for a few special $D$. Intermediate structures show a slip between two helices, keeping the relative positions of the other helices constant. Three different line-slips can thus arise, one for each of the three helices. As a result, a structure $(l,m,n)$ can line-slip into $(l\pm1,m\pm1,n)$, $(l\pm1,m,n\pm1)$ and $(l,m\pm1,n\mp1)$. Following Ref.~\cite{mughal2012}, we denote intermediate line-slip structures as $(l,m,\bm{n})$, $(l,\bm{m},n)$ and $(\bm{l},m,n)$, where the bold number identifies the helix that does not change during the transformation.

Note that in the regime $2.80\sigma\le D\le2.82\sigma$, a structure with particles along the cylinder core emerges, which is beyond the phyllotactic regime. We denote it instead by its point group symmetry, D$_5$. 

\section{Experimental Method}
\label{sec:expMeth}
The experimental assemblies considered here are obtained by sedimenting polystyrene particles into cylindrical pores on a polyvinyl alcohol (PVA) film. Instead of a compression rate, the tuning parameter is here the P\'eclet number $\mathrm{Pe}=\frac{v\cdot\sigma/2}{D_{\rm diff}}$, where $v$ is the sedimentation rate and $D_{\rm diff}$ is the self-diffusion constant.

To make the PVA film, we first spin coated AZ 9260 positive photoresist (AZ Electronic Materials USA Corp.) on a 3-inch silicon wafer (Addison Engineering, Inc.). The wafer was then soft baked at $110^{\circ}$C for 65 seconds and exposed to UV light (Karl Suss MA6/BA6) through a chrome-printed photomask (PhotoSciences, Inc.) with an array of well-defined circles of differing diameters at 13.5 mW/cm$^2$ for 140 seconds. Afterwards, the wafer was submerged in AZ 400K developer (AZ Electronic Materials USA Corp.), diluted 1:4 in deionized (DI) water for 180 seconds to form cylinders of crosslinked photoresist~\cite{shields2013field}, and then rinsed with DI water and dried with nitrogen gas. This last step was repeated until the uncrosslinked photoresist was completely removed. An SPTS Pegasus Deep Silicon Etcher was used to perform the deep reactive-ion etching (DRIE). The wafer was afterwards thoroughly cleaned to remove the photoresist. To prepare the PVA solution, 40g PVA powder (98\% hydrolyzed; Sigma-Aldrich, Co.) was dissolved in 400mL DI water at 200$^{\circ}$C, and 16mL glycerol (VWR International, LLC) was added to the solution. The PVA solution was then cast into a solid film as the negative mold of the etched wafer. The film was then gently removed from the wafer. The presence of glycerol rendered the film more elastic to help with the removal.

Monodisperse polystyrene particles of two different diameters (6.4$\mu$m, 7.7$\mu$m from Spherotech, Inc.) were used for sedimentation into the PVA film. The polystyrene particles were first re-suspended in a non-aqueous solvent mixture, consisting of ethanol, glycerol and Tween 20 (Sigma-Aldrich, Co.). Three different solvent mixtures were used, as listed in Table \ref{tbl:example}. To estimate the viscosity of mixtures, the Gambill method~\cite{gambill1959estimate} is used. Then $v$ and $D_{\rm diff}$ are calculated from Stoke's law and the Stoke-Einstein relation, respectively.

\begin{table}[h]
\small
  \caption{Non-aqueous solvent mixtures used to re-suspend polystyrene particles and properties of particles with $\sigma=7.7\mathrm{\mu}$m in these mixtures.}
  \label{tbl:example}
  \begin{tabular}{@{\extracolsep{\fill}}lllll}
    \hline
    Mix. & EtOH : Gly : Tween 20 & $v$/$\mathrm{\mu}$m$\cdot s^{-1}$ & $D_{\rm diff}/\mathrm{m}^2\cdot s^{-1}$ & Pe\\
    \hline
    1 & 1mL : 0.2mL : 1$\mu$L & 0.1 & $1\times10^{-15}$ & 400\\
    2 & 1mL : 0.4mL : 1$\mu$L & 0.03 & $0.5\times10^{-15}$ & 200\\
    3 & 1mL : 0.6mL : 1$\mu$L & 0.01 & $0.3\times10^{-15}$ & 100\\
    \hline
  \end{tabular}
\end{table}

A small segment was separated from the rest of the PVA film using a 6.00 mm biopsy punch (Ted Pella, Inc.), and the punched PVA film piece was placed into the lid of a 1.5mL Eppendorf tube. The re-suspended particle mixture was then pipetted in the lid directly above the PVA film.

To sinter particles, the segmented PVA film was first taken out of the suspension using metallic tweezers and then gently submerged in ethanol to wash off particles on the surface of the PVA film. The PVA film was placed on a glass slide, covered with a Petri dish to prevent dust accumulation and dried overnight at 25$^{\circ}$C. The glass slide was subsequently placed on a hot plate at 85$^{\circ}$C, covered with a Petri dish to sit for 20 minutes. After sintering the particles, the PVA film was dissolved by 0.5mL DI water at 75$^{\circ}$C in a glass scintillation vial (VWR International, LLC). The assembled fibrils were finally observed by scanning electron microscope (SEM) (FEI XL30 ESEM) with an accelerating voltage around 7kV.

Experiments with $D_1=2.25(10)\sigma$ and $D_2=2.70(10)\sigma$ were conducted. For the former, we find that Mixture 1 (see Table~\ref{tbl:example}) suffices to assemble ordered structures, even with such a large Pe, but for the latter a smaller Pe is needed. The height of the cylindrical pores mainly depends on the DRIE time and an average $\lambda_{\rm z}$ of 65$\rm\mu$m was observed. Experimental structures are matched to simulations by comparing the cylinder diameters and side-views alone. The measured $D/\sigma$ is first used to search for candidate structures, and then the phyllotactic indexes $l$, $m$ and $n$ are identified by visual inspection. Note, however, that the cylinder diameters are difficult to accurately determine due to edge effects of SEM imaging. Because more secondary electrons leave the sample at edges, rims of the pores appear brighter and are hard to pinpoint. The experimental $D$ was thus further calibrated by comparing with simulation predictions. The resulting error on $D$ then prevents the direct match of the experimental results to the phase boundaries obtained from the simulation sequences.

\section{Results and Discussion}
\label{sec:results}
In this section, we first present the equilibrium and assembly simulation results. We then relate these observations to the sedimentation experiments.

\subsection{Equilibrium Results}
\begin{figure}[ht]
\centering
  \includegraphics[width=1.0\columnwidth]{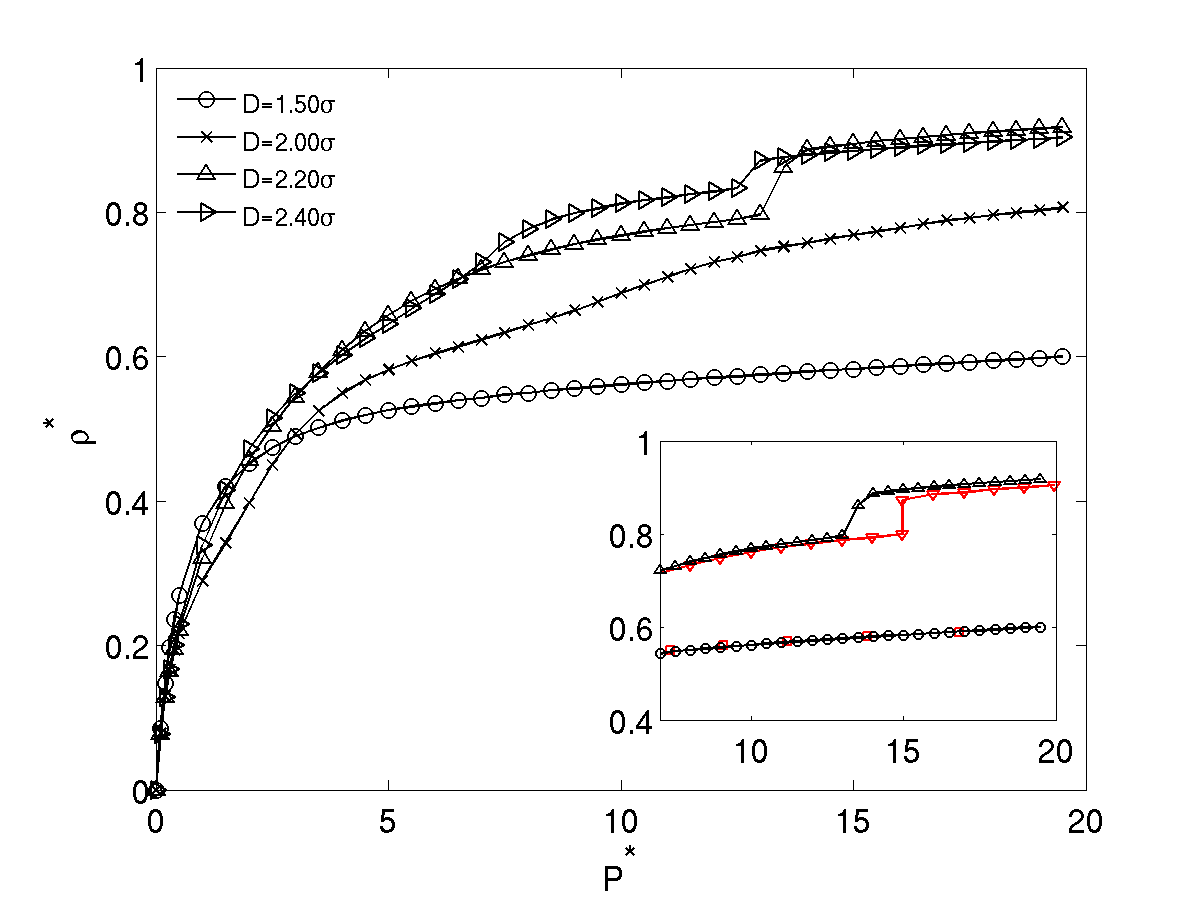}
  \caption{Equations of state for $D=1.50$, $2.00$, $2.20$ and $2.40\sigma$. The y axis is the reduced number density $\rho^*=N\sigma^3/V$. For $D<2\sigma$, no inflection point is observed in this pressure range. (Inset) Comparison with earlier results for $D=2.20\sigma$ (red downward triangles) and $D=1.50\sigma$ (red squares)~\cite{huang2009characterization} clearly show that sampling in the crossover regions for $D>2\sigma$ is more efficient.}
    \label{fig:eos}
\end{figure}

Qualitatively, our equations of state agree with earlier published results. For $D<2\sigma$ they are featureless, while for $D\ge 2\sigma$, inflection points are observed at intermediate pressures~\cite{duran2009ordering} (Fig.~\ref{fig:eos}). For $D<2\sigma$, complete quantitative agreement is also obtained, but for larger $D$, discrepancies are observed. For instance, at $D=2.20\sigma$ we observe the inflection point at $P^*=13.6$, while earlier work reported this feature at $P^*=15$~\cite{duran2009ordering,huang2009characterization} (Fig.~\ref{fig:eos} inset). Simulations with and without ECMC and torsional moves suggest that our enhanced sampling gives rise to a more robust equilibration in this hard-to-sample crossover regime.

\begin{figure}[ht]
\centering
  \includegraphics[width=1.0\columnwidth]{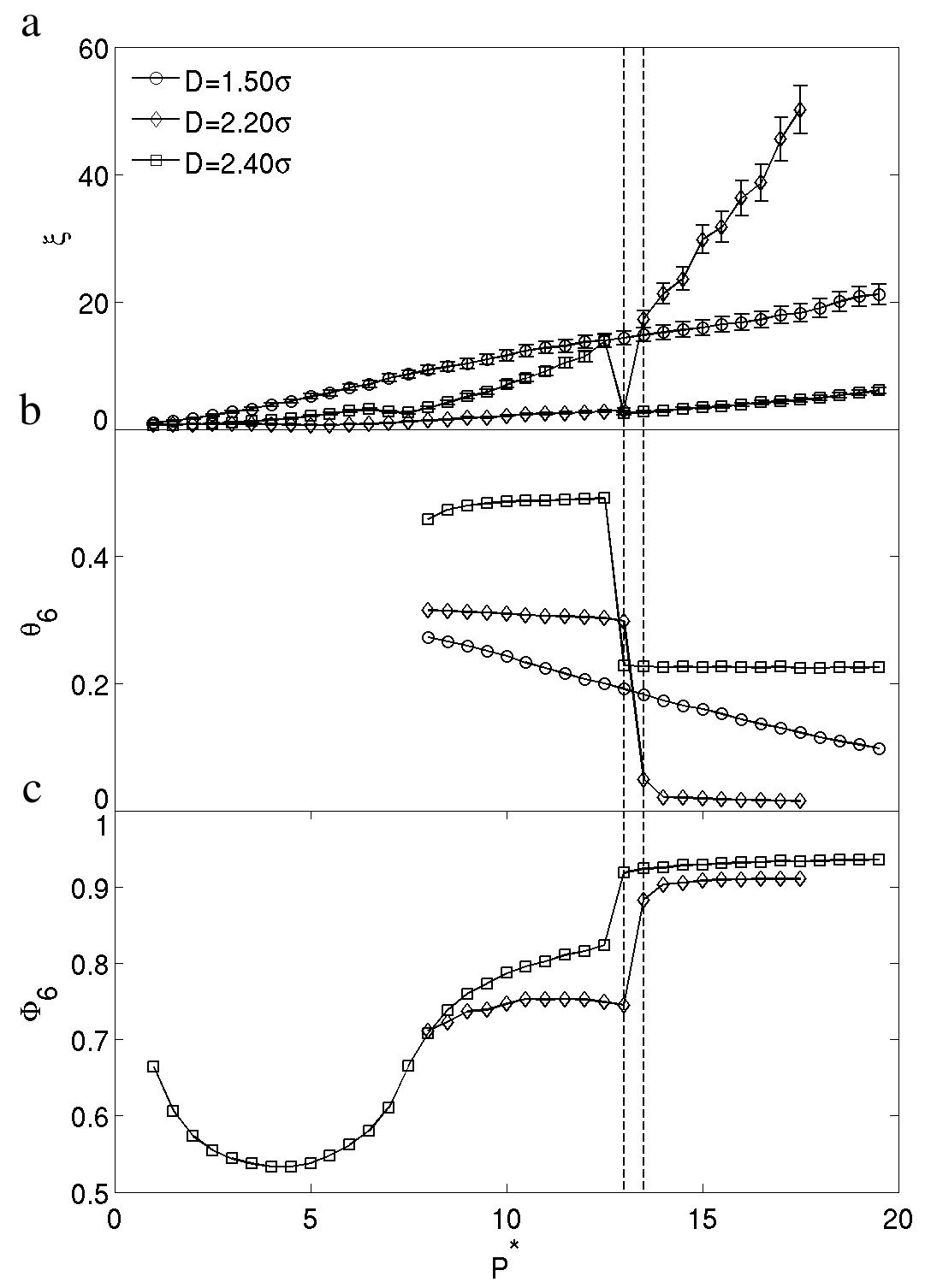}
  \caption{Pressure evolution of (a) the correlation length $\xi$, (b) the structural parameter $\theta_6$ and (c) the order parameter $\Phi_6$, for different $D$. As explained in the text, $\theta_6$ is not reported for disordered states, while $\Phi_6$ is not reported for $D<2\sigma$. Note that the sharp changes in the quantities coincide for different $D$ (dashed lines).}
    \label{fig:collTheory}
\end{figure}

Figure~\ref{fig:collTheory}a shows that the correlation length, $\xi$, generally increases with $P^*$. At low $D$, it grows monotonically (see, e.g., $D=1.50\sigma$ in Fig.~\ref{fig:collTheory}a), which is qualitatively similar to what happens at $D=\sigma$~\cite{lieb2013mathematical}. For $D>2\sigma$, however, the growth of $\xi$ is non-monotonic, and jumps in $\xi$ grow increasingly sharp with $D$. Interestingly, these non-monotonic regions accompany the inflection points in the equations of state. Although the jumps in $\xi$ are reminiscent of a phase transition, they are clearly not. The volume integral of the pair correlator is the susceptibility, $\chi$, which would diverge at a second-order phase transition, but here only peaks. Hence, even though phase transitions are formally impossible in a finite-pressure (quasi-)one-dimensional system with short-range interactions, clear crossovers between distinct structural regimes persist. The increasing sharpness of these jumps with $D$ is likely an echo of the first-order transition obtained in bulk three-dimensional systems, i.e., for $D\rightarrow\infty$.

\begin{figure}[ht]
\centering
  \includegraphics[width=1.0\columnwidth]{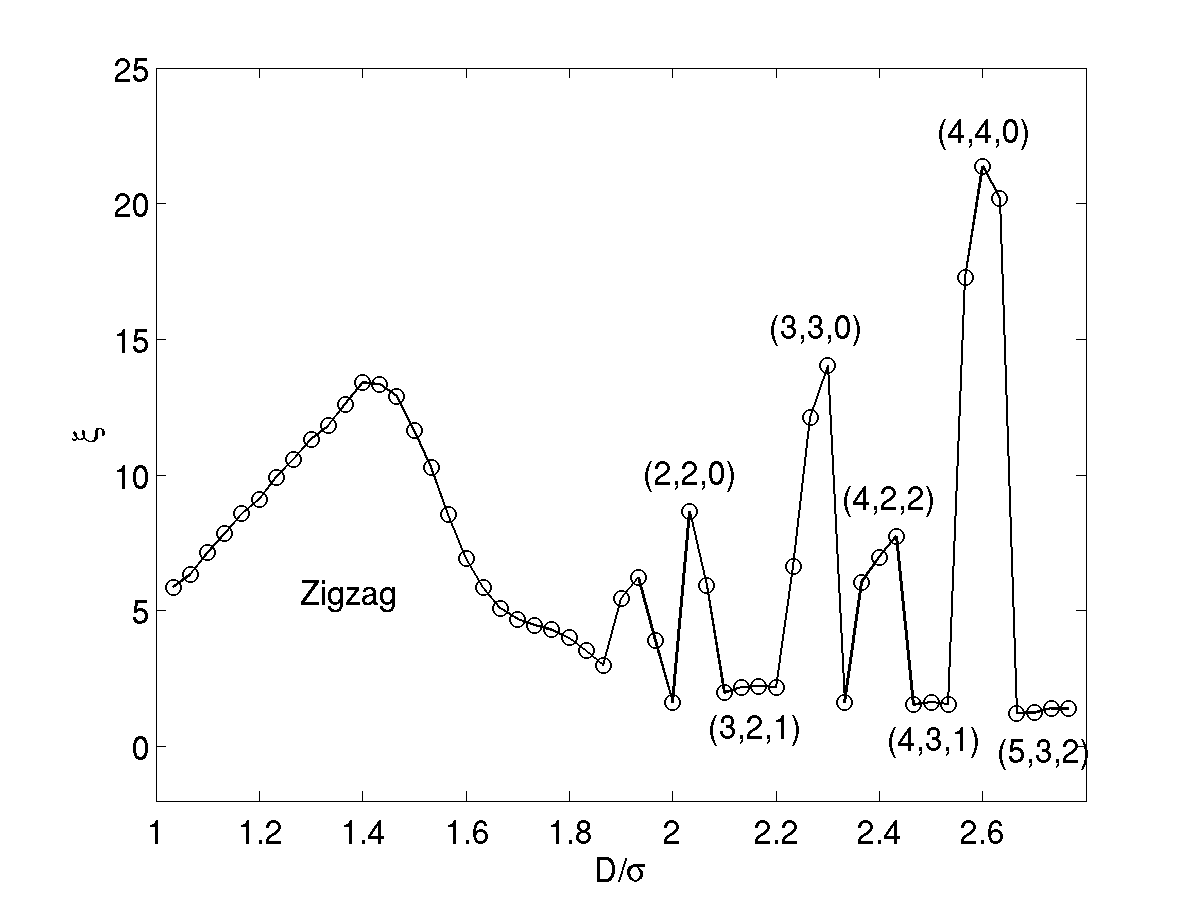}
  \caption{Structural correlation length, $\xi$, at $P^*=10$ for different $D$. For achiral structures with $D>2\sigma$, i.e., those with $(l,l,0)$ and $(l,l/2,l/2)$, $\xi$ is systematically larger than for chiral structures. Although $\xi$ is generally expected to grow with $D$, the rich set of structural intermediates gives rise to a non-monotonic behavior in this regime.}
    \label{fig:P10}
\end{figure}

The non-monotonic behavior of $\xi$ is accompanied by marked structural changes. For instance, the non-monotonicity of $\xi$ for $D=2.40\sigma$ corresponds to the system transitioning from an achiral double helix, i.e., (4,2,2), to a chiral single helix, i.e., (4,3,1). The systematic decrease of $\xi$ after the crossover reveals that as a new structural order develops its spatial extent momentarily shrinks. In other words, the increased packing efficiency enables the order to loosen. For achiral structures, however, the opposite is observed. For $(l,l,0)$ and $(l,l/2,l/2)$, $\xi$ grows much quicker than for chiral structures, and as a result $\xi$ increases (instead of decreasing) when the system goes from chiral to achiral (see, e.g., $D=2.20\sigma$ in Fig.~\ref{fig:collTheory}a). This effect may have a geometrical origin, and may also be an artifact of our definition of $\xi$. The projection of particle positions along $z$ indeed makes defects within the $x$-$y$ plane less prominent; planar rings hence appear more ordered than helices. The $D$ evolution of $\xi$ at $P^*=10$ clearly captures the net outcome (Fig.~\ref{fig:P10}). The correlation length displays multiple peaks, corresponding to distinct structures and to the systematic change in $\xi$ when a new order develops. Note that all the peaks for $D>2.00\sigma$ correspond to achiral structures.

In absence of genuine phase transitions, the structural parameter $\theta_6$ is expected to change continuously. Structural changes, however, are here so sharp that even with a resolution of $\Delta P^*=0.01$, the quantity appears to jump. We can thus confidently rely on it to distinguish different structural regimes (Figure~\ref{fig:collTheory}b). Remarkably, between structural crossovers, $\theta_6$ stays constant. Although continuous line-slip pathways between structures do exist, they do not seem to contribute significantly to the finite pressure equilibrium behavior. Otherwise, $\theta_6$ would change steadily. If line-slip structures do play a role at finite pressures, it is thus only very close to the structural crossovers. A clearer understanding of their effect is, however, beyond the resolution of our simulations. 

\begin{figure}[h]
\centering
  \includegraphics[width=1.0\columnwidth]{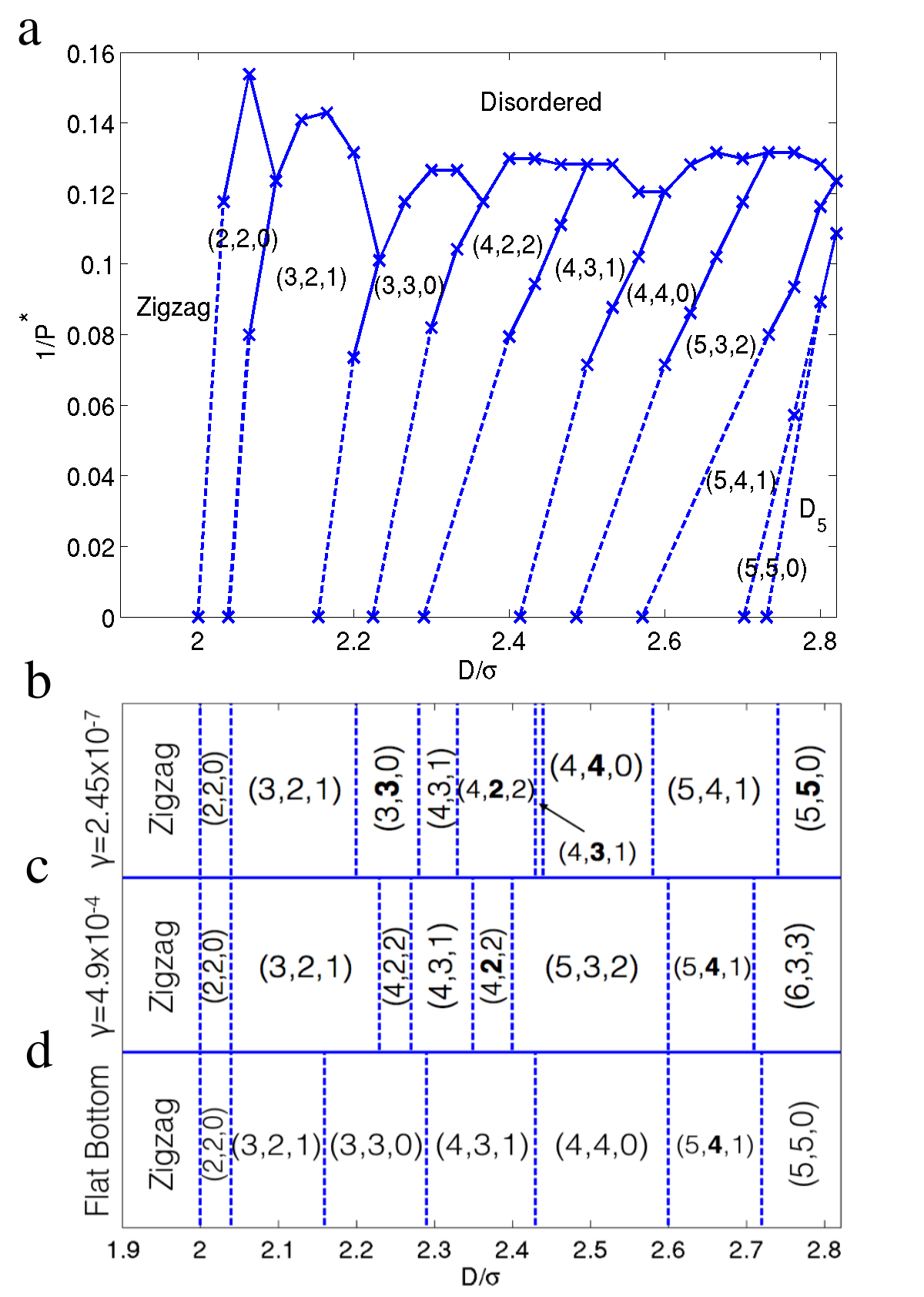}
  \caption{(a) Structure diagram for $\sigma<D\le2.82\sigma$. Because the zigzag regime extends all the way to $D=\sigma$, the $D<2\sigma$ region is featureless. Note that the disordered regime only has meaning for $D\geq2\sigma$, because no well-defined order-disorder crossover exists for $D<2\sigma$. The points at $P=\infty$ are the densest packing results from Ref.~\cite{mughal2012}. Finite pressure transition lines are obtained as described in the text. Structures that dominate the assembly of systems with (b) $\gamma=4.9\times10^{-4}$ and (c) $\gamma=2.45\times10^{-7}$ with periodic boundary conditions, and (d) $\gamma=2.45\times10^{-7}$ with hard boundaries. Differences between these the sequences are much more than just a shift in $D$. For instance, a new structure, (6,3,3), assembles only under fast compressions. Note that phase intervals labeled by non-line-slip structures also include the corresponding line-slip structures.}
    \label{fig:phases}
\end{figure}

As shown in Fig.~\ref{fig:collTheory}, all three quantities $\xi$, $\theta_6$ and $\Phi_6$, abruptly change together for $2\sigma<D\le2.82\sigma$. These changes also correspond to inflections in the equations of state. Figure~\ref{fig:phases}a summarizes these results as solid lines. The contrast with $D<2\sigma$, for which all quantities evolve smoothly and monotonically with pressure is marked. The distinction is reminiscent of the featureless equations of state for $D< 2\sigma$ and the simplicity of the corresponding zigzag order. 

For $2\sigma<D\le2.82\sigma$, the sequence of helical symmetries at finite $P^*$ systematically follows that observed in the densest packings.  Hence, starting from one state point, one obtains the same helical morphology by increasing (or decreasing) either $D$ or $P^*$. 
Although our simulation scheme is unable to equilibrate configurations in structural crossovers with $P^*\gtrsim 16$, we can use this observation to infer that the sequence of structures remains the same for $16\lesssim P^* <\infty$. Structural transitions in the densest packings at $P=\infty$~\cite{mughal2012,fu2016} can thus be used to extrapolate the equilibrium results (dashed lines in Fig.~\ref{fig:phases}). 
The correspondence between pressure and diameter is also physically interesting because the \emph{effective} interaction between finite-pressure hard spheres is not a step function at contact, but decays logarithmically with distance~\cite{brito2006rigidity,lerner2016emergent}. From the structural robustness, we conclude that (free-)energy minima in this regime are fairly insensitive to the precise form of the (effective) interaction form.
Crudely speaking, because the average gap between hard spheres scales as $\sim\frac{1}{P^{*}}$, spheres have an effective diameter $\sigma^*>\sigma$. Increasing pressure decreases the average interparticle gap and thus $\sigma^*$, hence giving rise to a comparable phase sequence as increasing $D/\sigma$ in this regime (Fig.~\ref{fig:phases}a).
Note, however, that the non-monotonic $\xi$ behavior for $D<2\sigma$ in Figure~\ref{fig:P10} is not observed in the pressure evolution of $\xi$ (see, e.g. $D=1.5\sigma$ in Figure~\ref{fig:collTheory}a), which hints that this correspondence may only apply within a specific diameter regime. 

At higher $D$, the structural complexity of the high-pressure packings also affects the robustness of the structural sequence. For instance, (5,5,0) does not appear between (5,4,1) and D$_5$ at $D=2.80\sigma$; (5,5,0) appears to be only stable at high pressures. The boundaries of (5,5,0) in Fig.~\ref{fig:phases}a are indeed all dashed. (Boundaries are here estimated by the results at $P^*=\infty$, and by knowing that (5,5,0) cannot be stabilized in the accessible pressure range for $P^*\lesssim16$.) The instability of (5,5,0) at higher $D$ and lower $P^*$ might be related to the ease of forming an inner core under thermal excitation, which naturally give rise to D$_5$. The optimal packings observed at higher $D$ might suffer a similar fate. 

\subsection{Simulation Assembly Results}
\label{sec:assembly}
Compressing a system sets a timescale, $\tau_\mathrm{comp}\sim \gamma^{-1}$, for equilibration. Because the equilibration timescale is itself determined by the timescale for structural rearrangement, $\tau_\alpha\sim\mathrm{e}^{\beta P\Delta V}$, where $\Delta V$ is an activation volume, a finite-rate compression is expected to trap the system in an intermediate structure around a pressure for which $\tau_\alpha\gtrsim\tau_{\mathrm{comp}}$. In a (quasi-)one-dimensional system, because $\Delta V$ is microscopic and because for $2\sigma<D\le2.82\sigma$ a cut through the phase diagram at fixed pressure is roughly equivalent to the high-density phase sequence, one might naively expect the assembly phase sequence to closely follow the densest packing sequence, only shifted in $D$. The results, however, reveal a much richer behavior.

Assuming $\Delta V\sim\sigma^3$ and knowing that at $P^*=16$ reaching equilibrium is computationally challenging, gives $\tau_\alpha\sim10^{7}$ as a typical timescale for a slow compression. For $\gamma=2.45\times10^{-7}$, we thus expect the assembly sequence to correspond to a cut through the phase diagram at $P^*\sim16$. Yet, a couple of discrepancies are observed (Fig.~\ref{fig:phases}b). Compared to the equilibrium results in Fig.~\ref{fig:phases}a, (4,3,1), which occurs after (4,2,2) at equilibrium, is here found between (3,3,0) and (4,2,2), while (5,3,2) is missing altogether. If the assembly sequence is a rough cut through the phase diagram, then it is not at a fixed pressure. 
In order to obtain a microscopic understanding of this effect, we consider the two outliers in more details. For $2.28\sigma\le D\le2.32\sigma$, even at such a slow compression, the crossover from (3,3,0) to (4,2,2) is not observed, suggesting that $\Delta V$ might be larger than for other transformations. Line-slips can continuously and easily transform one structure into another, hence it is natural to assume that structural crossovers should proceed via this route. The effect is indeed reminiscent of a martensitic transformation. It can also explain why $\Delta V$ is large for transforming (3,3,0) into (4,2,2): no single line-slip transformation between the two is possible. A line-slip through (3,\textbf{3},0) instead brings the system from (3,3,0) to (4,3,1). The same process accounts for the absence of (5,3,2). No single line-slip can transform (4,4,0) into (5,3,2), and thus (5,4,1) assembles even though (5,3,2) is the equilibrium structure. Additional evidence that $\Delta V$ is large for the (3,3,0) to (4,2,2)  crossover is that (4,\textbf{2},2) emerges around the point at which (3,3,0) disappears at equilibrium.

Because order develops slowly as pressure increases, fast compressions should lead to a cut through the phase diagram around the first structural crossover at $P^*\approx8$, which corresponds to a timescale $\tau_\alpha\lesssim5\times10^3$. Results for $\gamma=4.9\times10^{-4}$ indicate that structure skipping in this regime is in fact quite pronounced. For instance, (3,3,0) and (4,4,0) disappear. For $2.16\sigma\le D\le2.26\sigma$, (4,2,2) assembles because the system first gets trapped in (3,2,1), and then transforms into (4,2,2) through (3,\textbf{2},1). As a result (3,3,0) is  skipped. Similarly, (4,2,2) is skipped because (3,3,0) transforms into (4,3,1) via (3,\textbf{3},0). Other structures show the same effect, the most remarkable of which being the transformation of (5,3,2) via $(5,\textbf{3},2)$ into (6,3,3), which is \emph{not} a densest packing for any $D$. 
  
Surprisingly, (4,2,2) assembles even in regimes for which (3,2,1) does not appear in the structural diagram. It is key, however, to recall that $\Phi_6\ge0.7$ is chosen as a somewhat arbitrary criterion for ordering. Here, (3,2,1) does not appear in Fig.~\ref{fig:phases}a for $2.20\sigma<D\le2.26\sigma$, because then $\Phi_6\le0.7$ for that structure. Under such a quick compression, even the weak local order in the disordered regime, i.e., the liquid order, can thus affect assembly,  which is reminiscent of a geometrical frustration mechanism. As a result, a shift of the structural sequence to smaller $D$ is observed,  in contradiction to our initial expectation.  More importantly, this provides a new pathway for assembling dense packings in this regime. One can obtain the densest packing by quick compression, letting the system fall out of equilibrium at low densities and thus skip intermediate structure(s), rather than go through a slow crossover at high pressure.
 
\begin{figure}[h]
\centering
  \includegraphics[width=1.0\columnwidth]{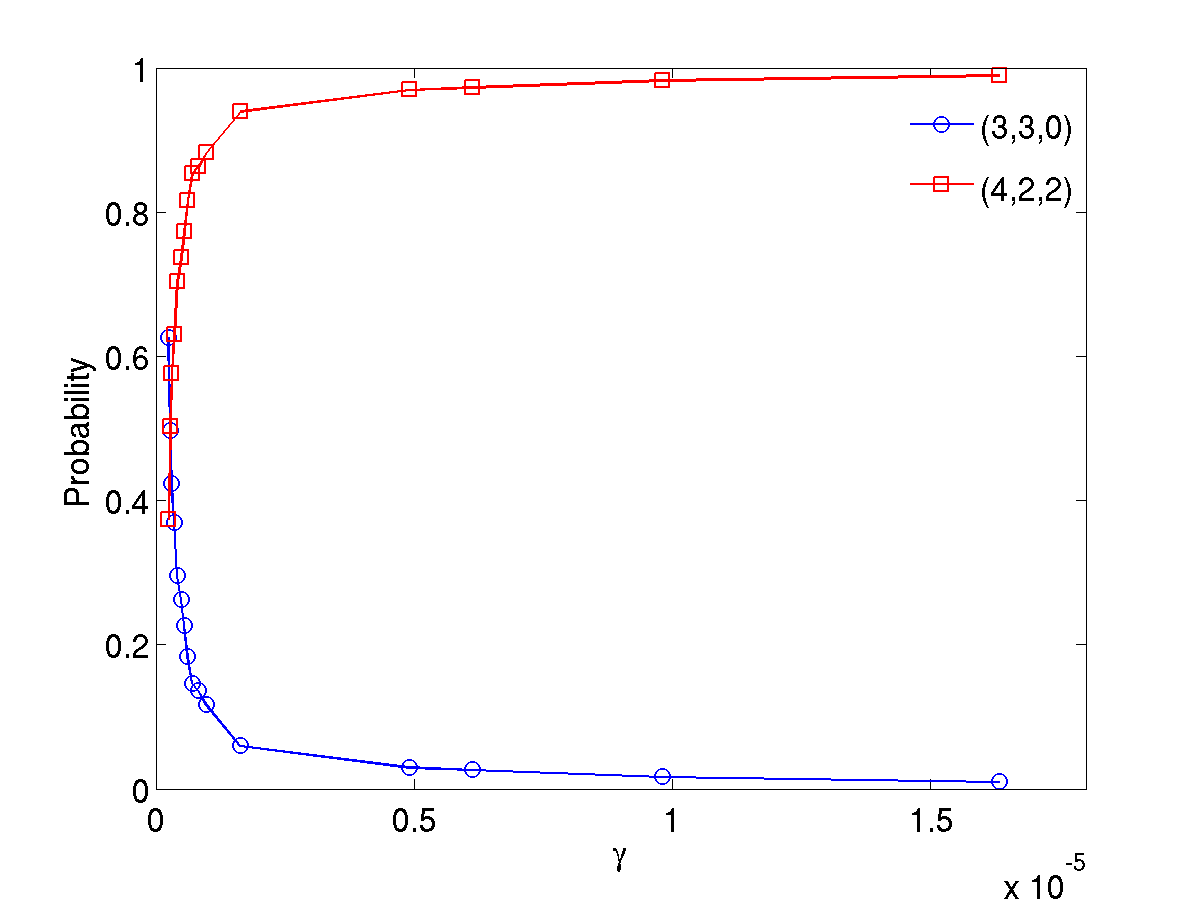}
  \caption{Probability of forming (3,3,0) and (4,2,2) at $D=2.20\sigma$ as a function of compression rate. Although (3,3,0) is the equilibrium structure, it is not as dynamically favorable as (4,2,2). As a result, the assembly probability of (3,3,0) decreases as $\gamma$ increases.}
    \label{fig:pvsv}
\end{figure}

The $\gamma$ dependence of the assembly sequence suggests that the compression rates select different crossover mechanisms. This dependence is a consequence of $\Delta V$ being different for each transition pathway. Generally, three line-slips are possible for each helical structure, and we expect the one with the smallest $\Delta V$ to be most favorable. If the dynamically favorable structure is not thermodynamically stable, however, its assembly probability should depend on $\gamma$. For example, for $D=2.20\sigma$, (3,2,1) forms at low pressures, while both (3,3,0) and (4,2,2) are accessible via line-slips ((4,3,1) simply does not fit within a $D=2.20\sigma$ cylinder). Figure~\ref{fig:pvsv} shows how the assembly probabilities for these two structures change with $\gamma$. Faster compressions clearly make (3,3,0) less likely and instead favor (4,2,2). The line-slip via (3,\textbf{2},1) has a smaller $\Delta V$, and hence is dynamically more favorable. To understand this effect, we analyze the above result with the densest packing sequence for $2.039\sigma\le D<2.1545\sigma$, i.e., (3,2,1), (3,\textbf{2},1) and (\textbf{3},2,1) (see Table 1 from Ref.~\cite{mughal2012}). We find that the dynamically favored line-slip ((3,\textbf{2},1) in this case) is the one that appears next to the current structure ((3,3,0) in this case) in the densest packing sequence. This mechanism, which also applies for other phyllotactic structures, reflects the importance of the packing efficiency for these transformations.

\begin{figure}[h]
\centering
  \includegraphics[width=1.0\columnwidth]{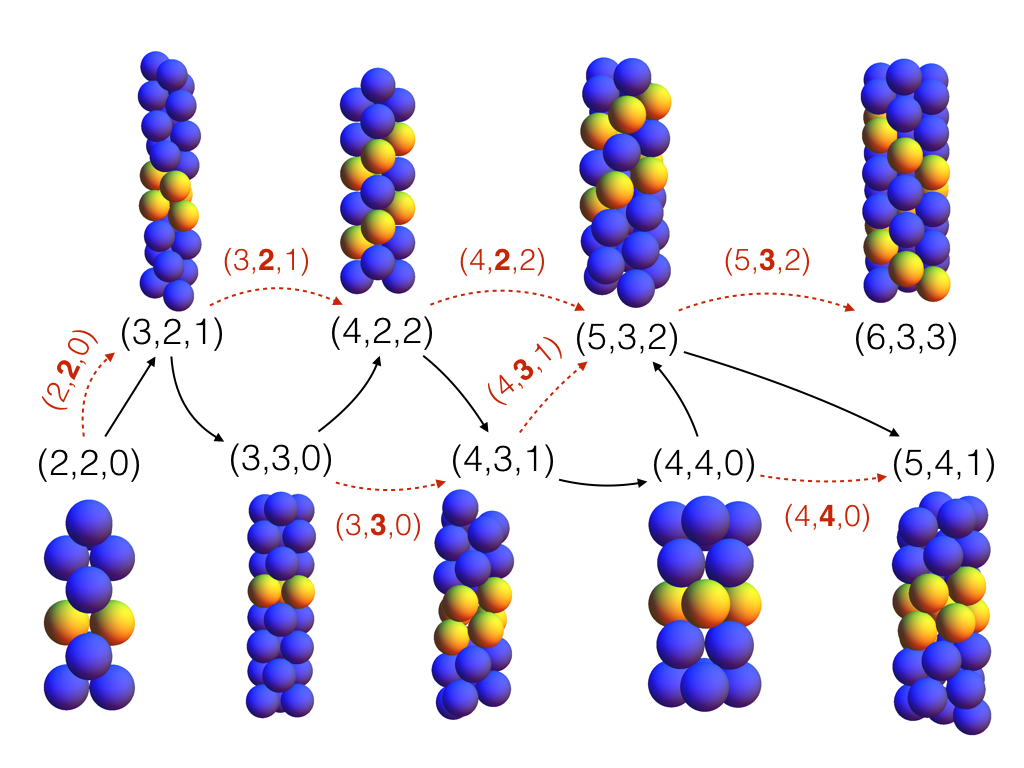}
  \caption{Schematic of the periodic configurations in the phyllotactic regime and their line-slip relationship. Black solid arrows indicate the assembly sequence and the red dotted arrows indicate the dynamically favorable pathways. Yellow particles are part of a same helix.}
    \label{fig:dynamics}
\end{figure}

Interestingly, of all the crossovers in the equilibrium structural diagram only passing from (2,2,0) to (3,2,1) is dynamically favorable. As a result, the boundary between these two structures is invariant of $\gamma$, while most of the other equilibrium structures at intermediate pressures can be skipped by fast compression. Figure~\ref{fig:dynamics} summarizes the dynamically favorable pathways by red dotted arrows. Note that no dotted arrow points to (3,3,0) and (4,4,0), so these two structures are not expected to assemble under quick compression.

All the simulation results mentioned above are for systems under periodic boundary conditions. In experiments, however, the cylinders are finite, with a hard bottom. Based on the results from the sequential deposition algorithm in cylinders~\cite{chan2011densest}, we expect this effect to play a role in assembly. We thus perform an additional series of fast compressions with hard boundaries (Fig.~\ref{fig:phases}d). The flat bottom clearly favors structures with planar staggered rings, $(l,l,0)$, which would otherwise not form under fast compression. In addition, between the $(l,l,0)$ regimes, only phyllotactic structures with $n=1$ assemble. The absence of (4,2,2), (5,3,2) and (6,3,3) likely results from their mechanical instability under these conditions.

\subsection{Experimental Realization}
\begin{figure}[h]
\centering
  \includegraphics[width=0.8\columnwidth]{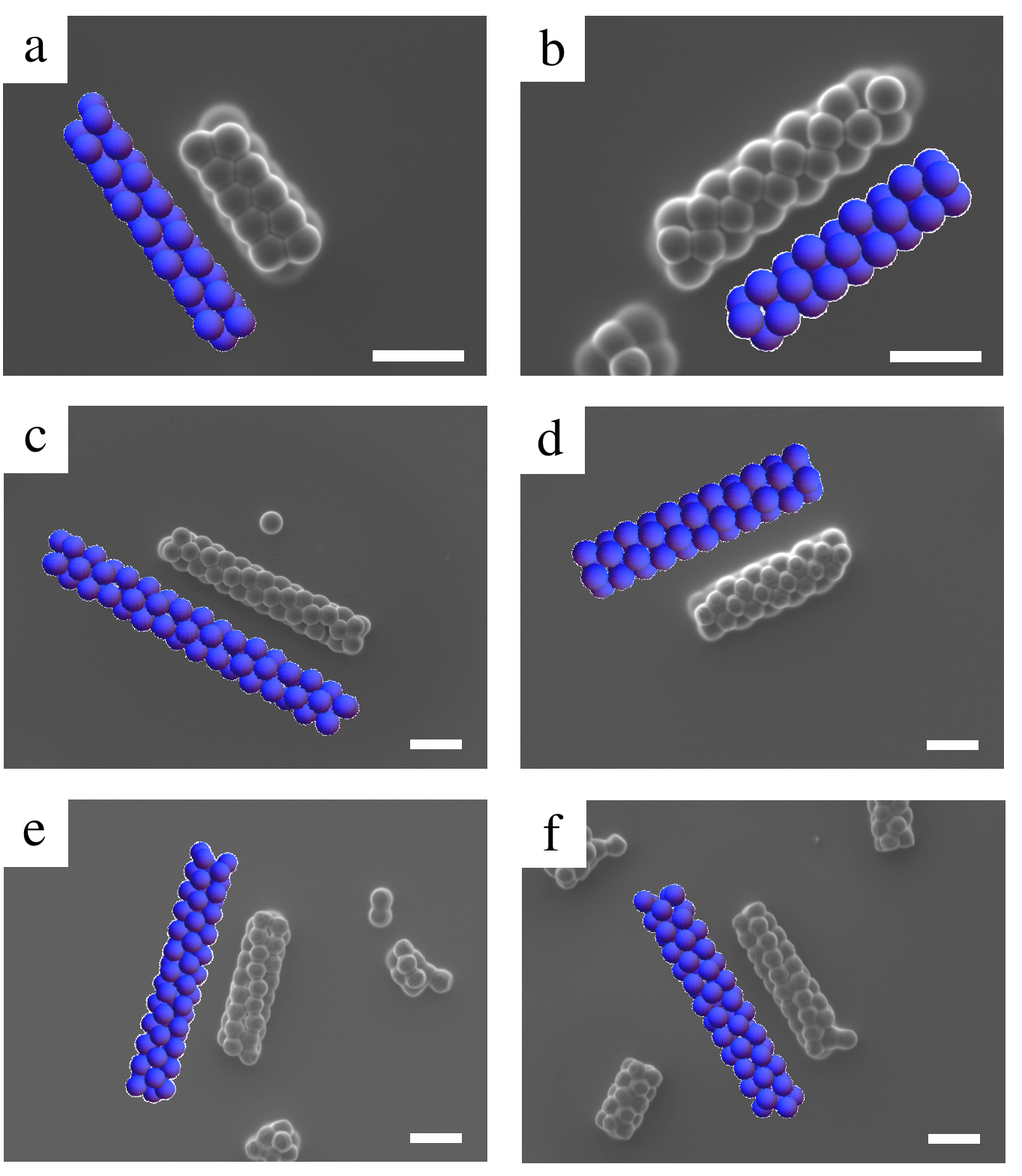}
  \caption{SEM images of structures assembled through sedimentation experiments. Grey spheres are polystyrene particles while blue spheres are simulation results with (a) (4,2,2), (b) (3,3,0), (c) (4,3,1), (d) (4,4,0), (e) (5,3,2) and (f) (5,4,1) structures. a, b and c have $D_1=2.25(10)\sigma$, while d, e and f have $D_2=2.70(10)\sigma$. Scale bars represent 20$\mathrm{\mu}$m.}
    \label{fig:simvsexp}
\end{figure}
To verify the generality and transferability of our model, we consider the assembly of micron-sized polystyrene particles in cylindrical pores by sedimentation. Calibrating the experimental $D$ from simulation predictions suggests that $D_1=2.25\sigma$ while $D_2=2.50\sigma$. As expected, different helical structures assemble at different $D$ (see Fig.~\ref{fig:simvsexp}). 
Experimentally, (3,3,0) and (4,4,0) dominate the assembly (Fig.~\ref{fig:simvsexp}b and d). Considering that the experimental $\lambda_{\rm z}\approx10\sigma$, assembly in a flat-bottom cylinder (Fig.~\ref{fig:phases}d) explains this observation. Yet some outliers from the simulation results are also found: (4,2,2) and (5,3,2) (Fig.~\ref{fig:simvsexp}a and e), most notably. Their presence in experiments might due to the density difference between particles and Mixture 3 being so small that the pressure at which these structures become mechanically unstable is not reached. Note that the assembly of (5,3,2) is a signature of fast compression, which is consistent with Pe>100 for the three experimental mixtures.

\section{Conclusions}

In this study, we have calculated the equilibrium structural behavior and out-of-equilibrium assembly of hard spheres confined to cylinders of diameter $\sigma<D\le2.82 \sigma$. At equilibrium, significant structural crossovers were identified, and for a given cylinder diameter the structural sequence upon changing $P^*$ and $D$ correspond. This suggests that the optimal packings are fairly robust to the effective interaction type in this diameter regime. Although the out-of-equilibrium assembly sequence depends on the compression (or sedimentation) rates and the choice of boundary conditions, 
the equilibrium behavior provides key insight into the results. Interestingly, mechanisms akin to geometrical frustration and martensitic transitions are found to play a role in the assembly process. 

Another interesting observation is that, although assembling densest packings for a given diameter can be difficult, the naive solution of following the equilibrium pathway may not always be optimal. For a system with multiple solid phases, one might aim for a continuous and diffusionless pathway connecting two solids, to avoid intermediate structures. 

In closing, we have here only considered assembly in fairly small cylinder diameters, but the phase morphology becomes increasingly complex as the diameter grows. One expects that even richer assembly processes might then be at play. 

\section{Acknowledgements}
We thank Crystal Owens, Christopher Reyes, Pearlson Prashanth, Joshua Socolar, Benjamin Wiley, and Sho Yaida for stimulating discussions. This work was supported by the National Science Foundation's (NSF) grant from the Nanomanufacturing Program (CMMI-1363483) and was performed in part at the Duke University Shared Materials Instrumentation Facility (SMIF), a member of the North Carolina Research Triangle Nanotechnology Network (RTNN), which is supported by the NSF (Grant ECCS-1542015) as part of the National Nanotechnology Coordinated Infrastructure (NNCI). CB was also supported by the China Scholarship Council.

\bibliography{rsc.bib}

\end{document}